\documentclass[sigconf,nonacm]{acmart}
\AtBeginDocument{%
  }

\usepackage{xcolor,soul,framed} %
\usepackage{xspace}
\colorlet{shadecolor}{yellow}

\usepackage{amsmath}
\usepackage{mdwmath}
\usepackage{mdwtab}
\usepackage{eqparbox}
\usepackage{url}
\usepackage{autobreak}
\usepackage{amsfonts} 
\usepackage{enumitem}

\def\eg{{e.g.,}\xspace}

\usepackage[font=small,skip=0.6pt]{caption}
\setcopyright{acmlicensed}
\acmBooktitle{UrbComp'25}

\newcommand{\name}{{\tt Judy}}


\begin{document}

\setlength{\abovedisplayskip}{3pt}
\setlength{\belowdisplayskip}{3pt}

\title{
Lessons from A Large Language Model-based  
Outdoor Trail Recommendation
Chatbot with 
Retrieval Augmented 
Generation
}

\author{Julia Ann Mathew and Suining He \\
\{julia.ann\_mathew,  suining.he\}\@uconn.edu \\
Ubiquitous \& Urban Computing Lab\\
School of Computing
\\
University of Connecticut
}

\begin{abstract}

The increasing popularity of outdoor recreational activities (such as hiking and biking) has 
boosted the demand for a conversational AI system to provide informative and personalized suggestion on outdoor trails. 
Challenges arise in response to
(1) how to provide accurate
outdoor trail information via conversational AI; and
(2) how to enable usable and efficient recommendation services. 
To address above, 
this paper discusses the preliminary and practical lessons learned 
from developing \name{}, an outdoor trail recommendation chatbot based on the large language model (LLM) with
retrieval augmented generation (RAG).  
To gain concrete system insights, we have performed case studies 
with the outdoor trails in Connecticut (CT), US. 
We have conducted web-based data collection, outdoor trail data management, and LLM model performance studies on the RAG-based recommendation. 
Our experimental results have demonstrated
the accuracy, effectiveness,
and usability of \name{}
in recommending outdoor trails based on the LLM with RAG. 
\end{abstract}


\settopmatter{printfolios=true}

\maketitle

\pagestyle{plain}

\section{Introduction}\label{sec:introduction}

The rising demand for the outdoor recreational activities has created the need 
for the outdoor trail recommendation~\cite{ivanova2023recommender}. 
Such recommendation aims to 
consolidate scattered knowledge such as trail location, difficulty, length, and permitted activities (e.g., biking), and provide accurate recommendation about trails for safe and pleasant activity planning.

To meet this need, this prototype study 
discusses the practical lessons and insights from the development 
of \name{},
a outdoor trail recommendation chatbot based on conversational AI. 
Toward development and deployment of \name{},
we have investigated the following
two questions. 
\textit{First}, how can we provide accurate
outdoor trail knowledge in response to complex user query?
Existing rule-based chatbots 
for recommendation largely follow predefined workflows. While they can handle specific and structured recommendation tasks, these chatbots could not handle complex or customized queries that deviate from their predefined rules, limiting their recommendation capability. 
General-purpose chatbots based on conversational AI, on the other hand, may lack the \textit{domain-specific knowledge} needed for effective outdoor trail recommendation. \textit{Second},  how can we enable usable and efficient recommendation services?
Existing outdoor trail recommendation platforms such as AllTrails~\cite{prah2024creating} offer static information. However, these platforms largely lack the \textit{conversational interactivity}. Therefore, the users often need to shift through multiple trail reviews to plan their trips. 

Toward addressing the above questions, we have performed real-world outdoor trail data studies (over 260 trails in Connecticut), and developed the system of \name{} as the large language model (LLM) recommendation chatbot with 
retrieval augmented generation (RAG)~\cite{lewis2020retrieval}.  
In particular, we have implemented RAG to retrieve 
the most relevant trail information 
from a constructed trail database, 
such that \name{} can provides accurate outdoor trail information. 
We have designed practical performance enhance mechanisms (such as caching)
to enhance retrieval efficiency, and enable usable recommendation services.  
We have performed 
real-world data analysis based on outdoor trail data in Connecticut (CT), US, and conducted
extensive experimental studies to validate the accuracy,
efficiency, and usability of \name{}
in recommending the outdoor trails.

The rest of the paper is organized as 
follows.
We first overview the system designs
in Sec.~\ref{sec:system}.
Then we present the design of \name{} with 
the LLM chatbot and RAG design in Sec.~\ref{sec:design}. 
We present the experimental 
studies and lessons learned in Sec.~\ref{sec:exp},
and conclude the paper in Sec.~\ref{sec:conclude}.

\begin{figure*}[!ht]
    \centering
    \includegraphics[width=0.88888\linewidth]{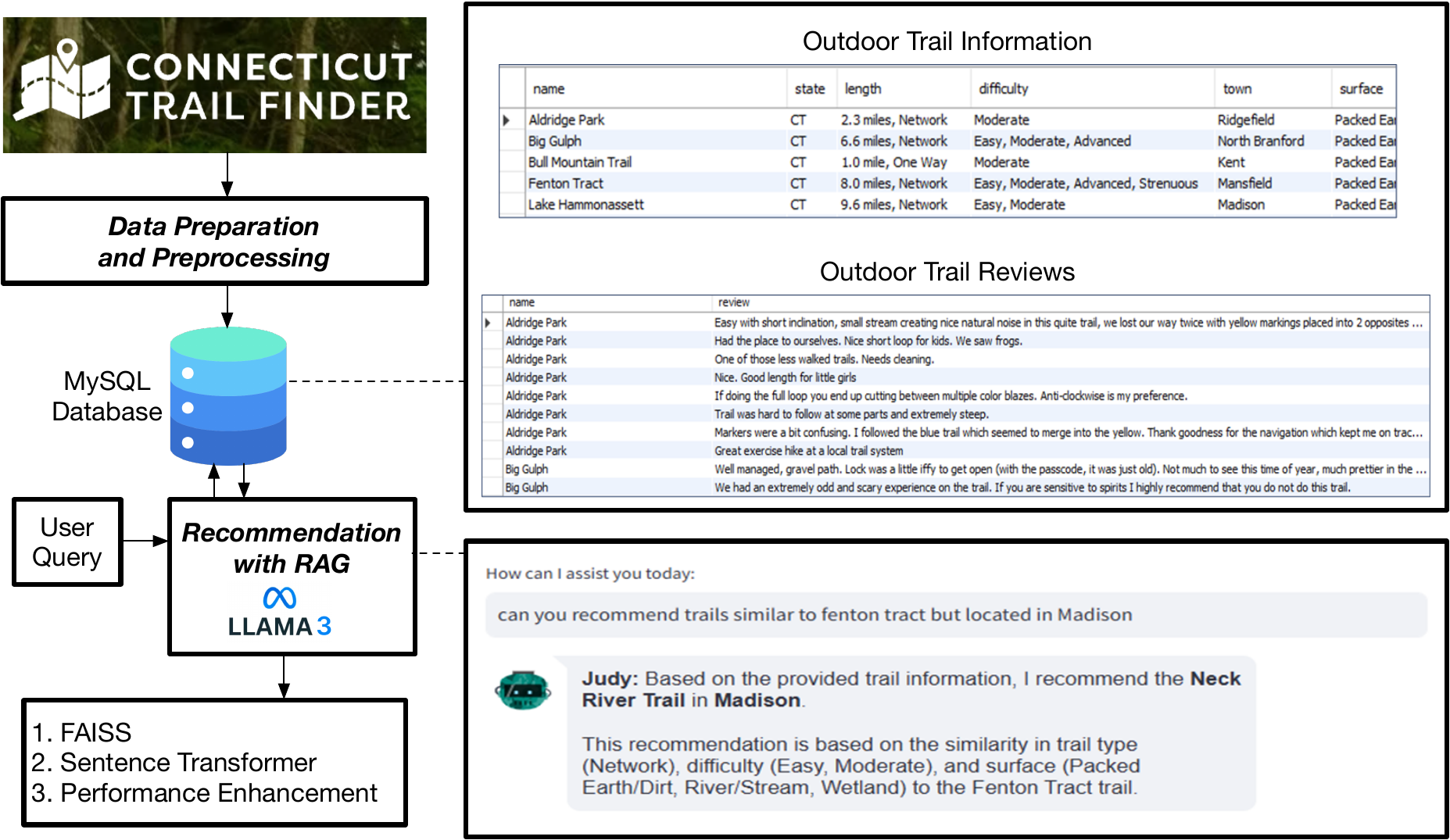}
    \vspace{0.06in}
    \caption{Overview of \name{} and samples of our outdoor trail information and collected reviews.}
    \label{fig:overview}
\end{figure*}

\section{System Overview}\label{sec:system}

The designs of \name{}, as showcased in Fig.~\ref{fig:overview},
consist of the following 
two phases:
(1) \textit{Data Preparation \& Preprocessing}:
In this phase, we pre-process and
prepare the basic outdoor trail 
information (including the trail name,
length, and difficult), and establish 
the MySQL database
hosted on the Amazon RDS
to streamline data access and the subsequent retrieval augmented generation. 
(2) \textit{User Query \& Recommendation
with RAG}:
Given the user query of the outdoor trail,
the LLM interprets the user query, 
and formulates a corresponding SQL query
that fetches the relevant data from the structured database.
The SQL query result, along with the user original question, are fed to the LLM.
The LLM leverages the above contexts to generate a structured and user-friendly response regarding the outdoor trail recommendation.

\section{Core Designs of \name{}}
\label{sec:design}

This section presents the core designs
of \name{}.
We first present the data preparation
and preprocessing in 
Sec.~\ref{subsec:datapre},
followed by the retrieval augmented 
generation designs in Sec.~\ref{subsec:rag}.

\subsection{Data Preparation 
and Preprocessing}\label{subsec:datapre}

In developing \name{},
we have crawled the basic information of outdoor trails
a credible platform called CT Trail Finder, such as length, difficulty, and pet permissions. 
To perform further natural language analysis, we have collected reviews for these trails from the Google Reviews and TrailLink~\cite{trail-link}. 
To prepare our analysis, we have performed data pre-processing. We have filtered away irrelevant trail reviews, removed emojis, excessive line spaces, asterisks, and other unnecessary characters. We have performed standardized abbreviations (such as replacing ``\textit{I've}'' with ``\textit{I have}'') to ensure the review consistency. We have re-structured the reviews such that  formatting is consistent across all reviews.

In our \name{} development, we have adopted 
the Selenium and BeautifulSoup for web scraping to gather trail-related data such as reviews, and trail features. To prepare the database, we have used MySQL, which is hosted on Amazon RDS, to form a relational database to store structured data, such as trail information and reviews.

\subsection{User Query and Recommendation with RAG}\label{subsec:rag}

The user queries in our current system designs typically pertain to features of the outdoor trail,
such as the trail’s length, 
location, difficulty level, 
permitted activities, 
trail accessibility, and other attributes.

The LLM will determine whether the user’s query is recommendation-relevant.
If yes, \name{} will leverage the LLM to generate a personalized response.
If the query is not recommendation-relevant, \name{} assesses whether it can be answered using the existing table schema. 
If the LLM in \name{} the user query can be resolved using the 
predefined schema, \name{} will perform the SQL query, and retrieve the necessary information regarding the outdoor trail.
If the user query requires more nuanced insights, such as opinions or user experiences, \name{} will invoke the RAG function to process the reviews of the relevant trail 
to construct an accurate and contextually rich response. Examples of such questions include 
``\textit{what do people say about the scenery on Aldridge trail}'', ``\textit{did visitors encounter wildlife at Big Gulph trail}'', 
``\textit{how crowded is the Pine Hill trail usually}'', 
and ``what do reviews say about biking on the Windsor Locks Canal trail''.

In terms of the RAG function,
\name{} will
retrieve the relevant reviews and their corresponding embeddings.
We have evaluated three kinds of 
sentence embeddings --- that is, 
Ollama (\texttt{nomic-} \texttt{embed-text})~\cite{hariprasath2024study}, and two kinds of pre-trained Sentence Transformers~\cite{devika2021deep}. 
One Sentence Transformer is 
trained question-answer (QA) pairs (\texttt{multi-qa-mpnet-base-cos-v1})~\cite{sentence-transformer2}, and maps sentences and paragraphs to a 512-dimensional dense vector space.
Another Sentence Transformer is the 
distilled version of multilingual Universal Sentence Encoder (\texttt{distiluse-base-multilingual-cased-v2})~\cite{cer2018universal,sentence-transformer3}. This transformer maps sentences and paragraphs to a 768-dimensional dense vector space and was designed for semantic search.

In our experimental studies,
we have observed that both the Sentence Transformers exhibited significantly faster response times compared to the Ollama embedding. 
Specifically, we have structured
and created a total of 
25 user queries (the same set of queries in Sec.~\ref{sec:design}) with the ground-truth answers for our testing evaluation. 
The Sentence Transformer pre-trained 
on QA pairs (\texttt{multi-} \texttt{qa-mpnet-base-cos-v1}) has delivered the fastest average response time at 0.73s, followed by 
2.89s by the distilled multilingual Universal Sentence Encoder 
(\texttt{distiluse-base-multilingual-cased-v2}),
and 4.69s by Ollama embeddings 
(\texttt{nomic-embed-text}).

In our \name{}, we adopt 
the Facebook AI Similarity Search (\texttt{FAISS})~\cite{douze2024faiss}
to rank the relevance of the reviews 
to the user's query. We find the 
top $k$ relevant reviews based on their similarity scores (cosine similarity based on a dot product on normalized embeddings).
These selected relevant reviews and the user’s query are fed to the LLM, and \name{} synthesizes them and generates a comprehensive and contextually appropriate responses.

In our \name{} development, we have 
used Llama3~\cite{touvron2023llama} for Judy’s conversational abilities, natural language understanding, and response generation. We have adopted LangChain to form the conversational flow and integrates Llama3 with MySQL for enhanced conversational functionality. We have adopted FAISS for similarity searches and nearest neighbor queries on trail reviews dataset for quick information retrieval.

In our experimental studies, we have observed 
the significantly long response time 
of using RAG from our outdoor trail database. 
To further enhance the efficiency of the RAG-based approach, we have implemented caching for the embeddings and reviews of the queried trails.

\section{Experiment and Preliminary Insights}
\label{sec:exp}

We find the measure of outdoor trail recommendation matching based on
the percentage of the correct answers against all the questions.
We measure the recommendation matching 
via the semantic similarities across the 
generated
responses and ground-truth answers
using the pre-trained sentence transformer (\texttt{multi-} \texttt{qa-MiniLM-L6-cos-v1})~\cite{sentence-transformer1}. 
We used and tested a total of 
25 user queries (the same set of queries in Sec.~\ref{sec:design}) with the ground-truth answers for our testing evaluation. 
We also compare \name{} with the same LLM backbone based on SQL only (without RAG).

\textit{Recommendation Matching}:
In our experiment, we have observed that \name{} achieved a recommendation matching of 96\%, which outperforms the 88\% accuracy of \name{} without RAG. 
\name{} with RAG retrieves the top 5 most relevant reviews to answer a user’s question. \name{} therefore provides concise and accurate responses. 
In contrast, \name{} without RAG needs to process all the review related to a specific outdoor trail through SQL, which can overwhelm the LLM and result in reduced response performance.

\begin{figure}
    \centering
    \includegraphics[width=\linewidth]{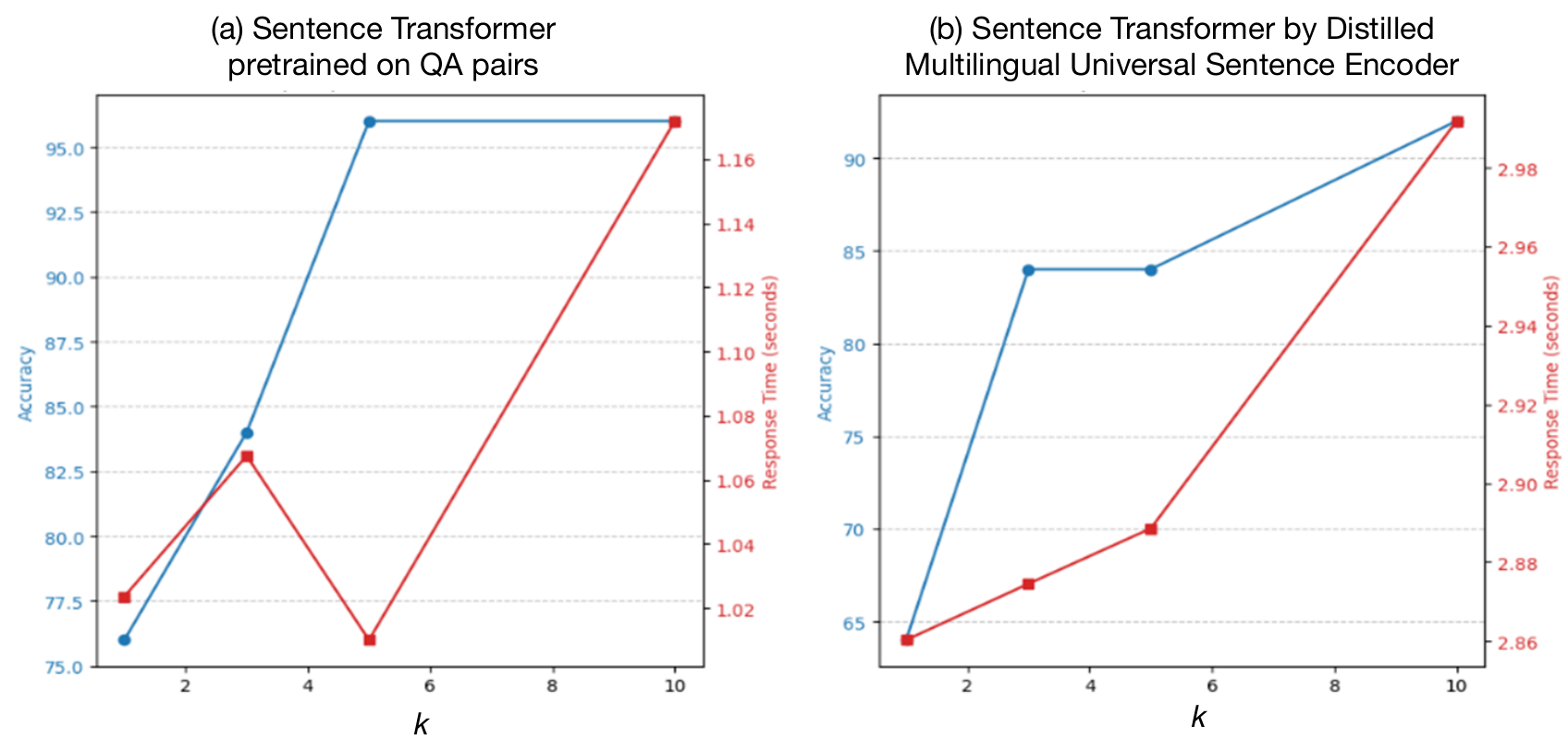}
    \caption{Sensitivity studies over $k$ for \name{} with two sentence transformers: (a) pretrained on QA pairs; and (b) distilled multilingual universal sentence encoder.}
    \label{fig:rag_k}
\end{figure}

\textit{RAG Designs}:
To determine the optimal value for k (the number of the outdoor trail reviews sent to the LLM backbone), we have evaluated impacts of different $k$'s on efficiency and recommendation matching.
In Fig.~\ref{fig:rag_k}(a),
we have shown the performance 
of \name{} with the embeddings 
using the sentence transformer 
pretrained on
QA pairs (\texttt{multi-qa-mpnet-base-cos-v1}).
We can see that $k=5$ yields an average response time of 0.73s and 
recommendation matching of 96\%.  We have 
also observed that when $k = 10$, the response time increases to 1.17 seconds, but the accuracy remains the same at 96\%. The increase in response time is due to processing of more trail reviews, which adds computational overhead without improving accuracy.

\section{Conclusion and Future Work}\label{sec:conclude}

We have developed \name{}, a large language model (LLM) based outdoor trail recommendation chatbot  based on 
retrieval augmented generation (RAG).  
We have performed data collection and
case studies 
with the outdoor trails in Connecticut, US. 
Our data analysis and experimental results have 
revealed the lessons learned 
in implementing RAG-based recommendation, and
demonstrated
the accuracy, effectiveness,
and usability of our chatbot
in recommending outdoor trails.

In our future studies, we will 
perform more in-depth studies on:
(1) integrating more data sources (\eg weather, social network, city planning~\cite{10.1145/3589132.3625588,9363542,han2024micromobility})
to enhance the outdoor trail recommendation
and usability;
and (2) involving more user studies
to gain user-centered insights
into their receptivity and acceptability 
about \name{}.

\section*{Acknowledgment}
This project is supported, in part, by the National Science Foundation (NSF) under Grants 2303575 and 2239897, and the Connecticut Division of Emergency Management \& Homeland Security
(DEMHS) Hazard Mitigation Grant Program (HMGP).
Any opinions, findings, and conclusions or recommendations expressed in this material are those of the authors and do not necessarily reflect the views of the funding agencies.

\bibliographystyle{ACM-Reference-Format}
\bibliography{acmart.bib}

\end{document}